\documentstyle[11pt]{article}  
\input{epsf}

\def\spacing #1{\small \renewcommand{\baselinestretch}{#1} \normalsize}

\newcounter{nr}
\newcounter{subnr}[nr]
\newcounter{subsubnr}[subnr]
\newcommand{\nsection}[1]{\stepcounter{nr} \par \vspace{2ex}
  \begin{center} {\thenr.~~#1} \end{center}
  \nopagebreak \vspace*{-1.3ex}}

\def\thebibliography#1{\par \vspace{2ex}
 \begin{center} {REFERENCES} \end{center}
 \nopagebreak \vspace*{-1.3ex} \list
 {\arabic{enumi}.}{\settowidth\labelwidth{3ex}\leftmargin\labelwidth
 \advance\leftmargin\labelsep
 \usecounter{enumi}}
 \def\newblock{\hskip .11em plus .33em minus -.07em}
 \sloppy
 \sfcode`\.=1000\relax}

\begin{document}
\spacing{0.9091}
\parskip=0ex \parindent=0ex
\small

\twocolumn[{%
\begin{center}
    {\large \bf  MODE TYPING 
    OF THE DOMINANT MODE IN THE DELTA SCUTI STAR X CAE BY 
     LINE PROFILE MOMENTS}\\[4ex]   
    {\normalsize \bf
L. Mantegazza$^{1,2}$, E. Poretti$^1$}\\[4ex]
    \begin{minipage}[t]{16cm}
      \begin{tabbing}               
        $^1$ Osservatorio Astronomico di Brera, Via Bianchi 46,
        I--22055 Merate, Italy\\
        $^2$ Dip. Fisica Nucleare e Teorica, Universit\a`a di Pavia, Italy
      \end{tabbing}
    \end{minipage}
\end{center}}]

\begin{center} {ABSTRACT} \end{center} \vspace*{-1.3ex} \parskip=1ex
We use moment method developed by Balona (1987) to identify the dominant 
mode of the multimode pulsating $\delta$ Scuti star X Cae.
It is shown as a $\ell=1$ and $m=-1$ non-radial mode gives a satisfactory
fit of the amplitudes and phases of light and line moments variations
provided that the rotational axis has $i\geq 45^o$.
The physical parameters corresponding to this mode are derived and 
discussed.


\nsection{INTRODUCTION}

One fundamental step in order to perform asteroseismology with $\delta$ Scuti 
stars is the typing of their pulsation modes. The recent multisite campaigns
have shown that these objects have very rich pulsational spectra, however
the photometric studies, while allow accurate mode detections, are inadequate
to identify them. One powerful tool to this end is the study of line profile
variations which are suitable to be analyzed with different techniques.
In particular the technique of analysis of the low order moments
developed by Balona (1986a, 1986b, 1987b, 1996)
is very promising for the study of low order non radial modes, which are
those responsible of light variations.

In this paper we will show the result supplied by it to the study of
the dominant mode of the star X Cae. Our photometric and spectroscopic studies
 of this star 
(Mantegazza and Poretti, 1992 (PaperI), 1996 (PaperII))  have
shown the presence of at least 14 pulsation modes with one of them far
stronger than the others ($\nu=7.393$ c/d, $A_B=0.049$ mag; all the others
have amplitudes below 0.01 mag). 
A preliminary analysis of the line moments has
been already performed in Paper II; here we present an improved
analysis applied to the moments of the dominant mode, made following 
Balona et al. (1996).
All the data relevant for this analysis are in Paper II.
We remind that spectra with 50000 resolution were obtained at 
La Silla Observatory with the CAT telescope on November 26--29, 1992.

In the meantime, with the availability of the Hipparcos parallaxes (Antonello,
private communication), it has been possible to fix the stellar absolute
magnitude, which has been matter of discussion in Paper II since values
obtained from different $uvby\beta$ calibrations were in partial disagreement.
The value obtained from Hipparcos data is $M_V=1.27\pm0.15$ which agrees
with that derived from Crawford (1979) calibration ($1.05\pm0.30$) in Paper II.
The $ubvy\beta$ photometry allows to derive also $T_{\rm eff}=7000^o$K,
so by combining it with the Hipparcos absolute magnitude we get $R=3.5R_{\odot}$.
\nsection{APPLICATION OF THE METHOD OF MOMENTS}

The Balona's method assumes that the pulsation velocity amplitude is much 
smaller
than the projected rotational velocity and that the ratio between rotation
and pulsation frequencies ($\Omega/\omega$) is small. These conditions 
are both verified by the dominant mode of X Cae
which has a radial velocity amplitude of 3.6 km/s
to be compared with $v_e \sin i=70$~km/s, and, by combining this with the 
radius, we get $P_{rot}/\sin i=2.55$d and thus
 $\Omega/\omega\geq 0.05$.

The amplitudes and phases of line moment variations and light variations
can be expressed in terms of three velocity fields $v_r$ (vertical velocity),
$v_h$ (horizontal velocity) and $v_f$. This last is a fictitious velocity 
field which quantifies the distortion of the line profiles generated by
the variation of the temperature distribution on the photosphere due to the
non--radial pulsations and is defined as $v_f=\Delta F/F\cdot v_e\sin i$,
where $F$ is the flux. 
For each possible non--radial mode (identified by $\ell$ and $m$ ) and
a particular inclination of the rotational axis $i$ the amplitudes and
phases of the light curve and of the first few moments can be fitted to the
predicted ones and the pulsational parameters computed.
Since there are 6 unknowns (amplitudes and phases of the three velocity 
fields)
by using at least three moments and the
light curve, it is possible to solve the system with the least squares,
and in this case the residual defines a {\it discriminant} that allows
to compare solutions with different ($\ell,m,i$) sets and obtain the 
most likely ones. The discriminat has dimension of velocity since all
the equations of condition in the algorith are normalized to velocities.

Since the moment accuracy decreases as the order increases only the first few
moments can be used, so in order to have a significant discriminant
it is useful to employ as many constraints as possible.
In the case of $\delta$ Scuti stars $v_h< 0.08\cdot v_r$ (
since $v_h\simeq74.4\cdot Q^2\cdot v_r$, $Q$ pulsational constant),
so that it is reasonable to constrain $v_h=0.08v_r$ and with the same phase,
and so to reduce the unknown velocity fields to two.
Another constraint is that $v_f<v_e \sin i$, since larger values imply relative
flux variations greater than unity.

For our application to X Cae we have used amplitudes and phases of the first 5
moments and of the $B$ light curve (this last has been chosen because the lines
used to compute the moments are at about $4500$\AA).
Light and moments equations have been weighed according to the standard 
errors of their amplitudes and phases as obtained from the Fourier fit.
The computation included all the modes with to $0\leq \ell\leq4$
and $-\ell\leq m \leq \ell$.
The discriminant is plotted againts angle of inclination in Fig.1 for
the modes which best fit the data.

\begin{figure}[htbp]
\epsfxsize=7cm
\centerline{\epsfbox[29 19 550 780]{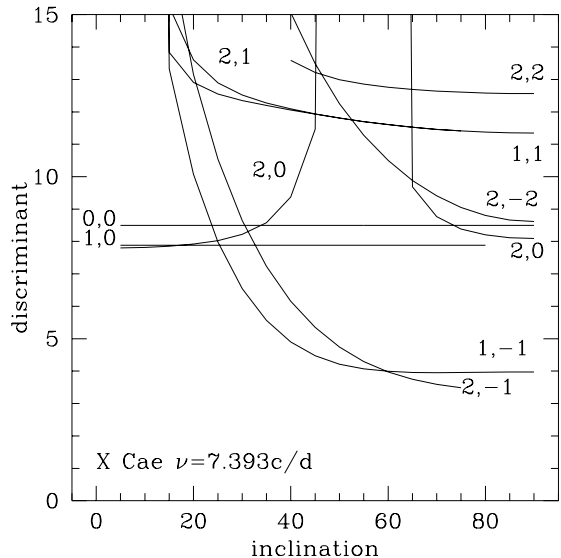}}
    \caption[]{\label{fig:myfig}	
Discriminant (arbitrary units) as a function of angle of inclination
(in degrees) for various modes ($\ell,m$)}			
\end{figure}
\nsection{DISCUSSION}
It is apparent that ($1,-1$) and ($2,-1$) are the best candidates.
Both are prograde modes ($m<0$) and this is in agreement with a visual
inspection of the line profiles which show that the dominant perturbation
is moving in the prograde direction (see Fig.5 of Paper II).
Moreover if we look at the velocity amplitudes  we see that for ($2,-1$)
as its discriminant approaches its minimum values $v_f$ approaches $v_e\sin i$
and overcomes it for $i>75^o$, so, while this mode supply a good 
mathematical fit of the observed quantities, it is not physically meaningful.
Therefore the most promising identification for
$\nu=7.393$ c/d is ($1,-1$). This mode matches almost perfectly amplitude
and phase of the light curve. The minimum discriminant is attained at 
$i=70^o$ and for this inclination $v_r=19$~km/s and $v_f=14~$km/s,
this velocity implies an amplitude of the temperature variation of about
$300^o$K. The ratio between flux and radius variations results $f=15$
and the phase difference between these two quantities is $\psi=118^o$
and $\Delta R/R=0.013$. As a comparison, the Cepheids, which are radial 
pulsators, have $<\psi>=124^o\pm 12$ and 
$5\leq f \leq 15$ (Balona \& Stobie, 1979).

Obviously we can see from the figure that the inclination is not strongly 
constrained and any value with $i>45^o$ it could be acceptable. In this range
$18<v_r<24$, and $13<v_f<18$.
In order to fully constrain $i$ could be useful to get similar solutions
for the other observed modes. Unfortunately the present data do not allow
to estimate amplitudes and phases of their moments with the required
accuracies. To this end a new set of spectroscopic data covering a 
longer time baseline is necessary.
A 8~--~day run is scheduled on November 1996 always at the ESO--CAT telescope.





    



\begin{thebibliography}{1}

\bibitem{1}
Balona L.A., 1986a MNRAS 219, 111
\bibitem{2}
Balona L.A., 1986b MNRAS 220, 647
\bibitem{3}
Balona L.A., 1987  MNRAS 224,41
\bibitem{4}
Balona L.A., Stobie R.S., 1979 MNRAS 189, 649
\bibitem{4}
Balona L.A. et al., 1996  MNRAS 281, 1315
\bibitem{5}
Crawford D.L., 1975 AJ 80, 955
\bibitem{6}
Mantegazza L., Poretti E., 1992 A\&A 255, 153 
\bibitem{7}
Mantegazza L., Poretti E., 1996 A\&A 312, 855 
\smallskip

\end{thebibliography}
\end{document}